\documentclass[a4paper,fleqn]{cas-dc}

\usepackage[round,authoryear,sort&compress]{natbib}

\usepackage{amsmath,amssymb,amsfonts}
\usepackage{algorithm}
\usepackage{graphicx}
\usepackage{textcomp}
\usepackage{stfloats}
\usepackage{url}
\usepackage{verbatim}
\usepackage{caption}
\usepackage{algpseudocode}
\usepackage{pifont}
\usepackage{multirow}
\usepackage{booktabs}
\usepackage{svg}

\def\tsc#1{\csdef{#1}{\textsc{\lowercase{#1}}\xspace}}
\tsc{WGM}
\tsc{QE}
\tsc{EP}
\tsc{PMS}
\tsc{BEC}
\tsc{DE}

\begin{document}
\let\WriteBookmarks\relax
\def\floatpagepagefraction{1}
\def\textpagefraction{.001}
\shorttitle{Multimodal Medical Endoscopic Image Analysis via Progressive Disentangle-aware Contrastive Learning}
\shortauthors{Junhao Wu et~al.}

\title [mode = title]{Multimodal Medical Endoscopic Image Analysis via Progressive Disentangle-aware Contrastive Learning}                      


\author[1]{Junhao Wu}
\ead{wujunhao333@gmail.com}

\author[2]{Yun Li}

\author[1]{Junhao Li}

\author[1]{Jingliang Bian}

\author[3]{Xiaomao Fan}

\author[2]{Wenbin Lei}
\cormark[1]

\author[1]{Ruxin Wang}
\ead{rx.wang@siat.ac.cn}
\cormark[1]


\cortext[1]{Corresponding author.}
\tnotetext[0]{Junhao Wu, Yun Li, and Junhao Li contributed equally.}


\affiliation[1]{organization={Shenzhen Institutes of Advanced Technology, Chinese Academy of Sciences},
                city={Shenzhen},
                state={Guangdong},
                postcode={518055}, 
                country={China}}

\affiliation[2]{organization={First Affiliated Hospital, Sun Yat-sen University},
                city={Guangzhou},
                state={Guangdong},
                postcode={510080}, 
                country={China}}

\affiliation[3]{organization={College of Big Data and Internet, Shenzhen Technology University},
                city={Shenzhen},
                state={Guangdong}, 
                postcode={518118}, 
                country={China}}

\begin{abstract}
Accurate segmentation of laryngo-pharyngeal tumors is crucial for precise diagnosis and effective treatment planning. However, traditional single-modality imaging methods often fall short of capturing the complex anatomical and pathological features of these tumors. In this study, we present an innovative multi-modality representation learning framework based on the `Align-Disentangle-Fusion' mechanism that seamlessly integrates 2D White Light Imaging (WLI) and Narrow Band Imaging (NBI) pairs to enhance segmentation performance. A cornerstone of our approach is multi-scale distribution alignment, which mitigates modality discrepancies by aligning features across multiple transformer layers. Furthermore, a progressive feature disentanglement strategy is developed with the designed preliminary disentanglement and disentangle-aware contrastive learning to effectively separate modality-specific and shared features, enabling robust multimodal contrastive learning and efficient semantic fusion. Comprehensive experiments on multiple datasets demonstrate that our method consistently outperforms state-of-the-art approaches, achieving superior accuracy across diverse real clinical scenarios.
\end{abstract}




\maketitle

\section{Introduction}
Laryngeal cancer is a common malignant tumor in the head and neck region, posing significant burdens on public healthcare service~\citep{steuer2017update}. According to GLOBOCAN, China reported $29,500$ new cases and $16,900$ deaths from laryngeal cancer in 2022~{bray2024global}. Accurately delineating laryngo-pharyngeal tumor region is crucial for differentiating pathological tissues from healthy ones, minimizing unintended damage during interventions, and optimizing therapeutic outcomes such as radiotherapy dose delivery~\citep{oreiller2022head,huang20213d}.

Endoscopic imaging, owing to its high clarity and resolution, along with realistic colors, serves as a crucial technique for the early screening of cavity neoplasm~\citep{subramanian2014advanced}. Two widely used endoscopic imaging modalities are White Light Imaging (WLI) and Narrow Band Imaging (NBI). WLI provides detailed morphological information, capturing macroscopic features such as tissue texture, color, and boundary structures. NBI, on the other hand, enhances the visibility of superficial vascular structures by exploiting hemoglobin’s specific absorption characteristics under narrow-band blue and green light, making it particularly useful for detecting microvascular changes indicative of early neoplastic transformations~\citep{yumii2024laryngopharyngeal, paderno2022videomics}. Despite the individual strengths of these modalities, relying on a single imaging source often fails to capture the full spectrum of tumor characteristics. WLI’s ability to detail tissue morphology is offset by its limited sensitivity to early vascular changes, while NBI excels in enhancing vascular patterns but provides poor structural context and is more susceptible to noise and low contrast~\citep{teevno2024domain}. Consequently, single-modality approaches often struggle to achieve the high segmentation accuracy and robustness required in complex clinical scenarios.

\begin{figure}[t]
    \centering
    \includegraphics[width=\columnwidth]{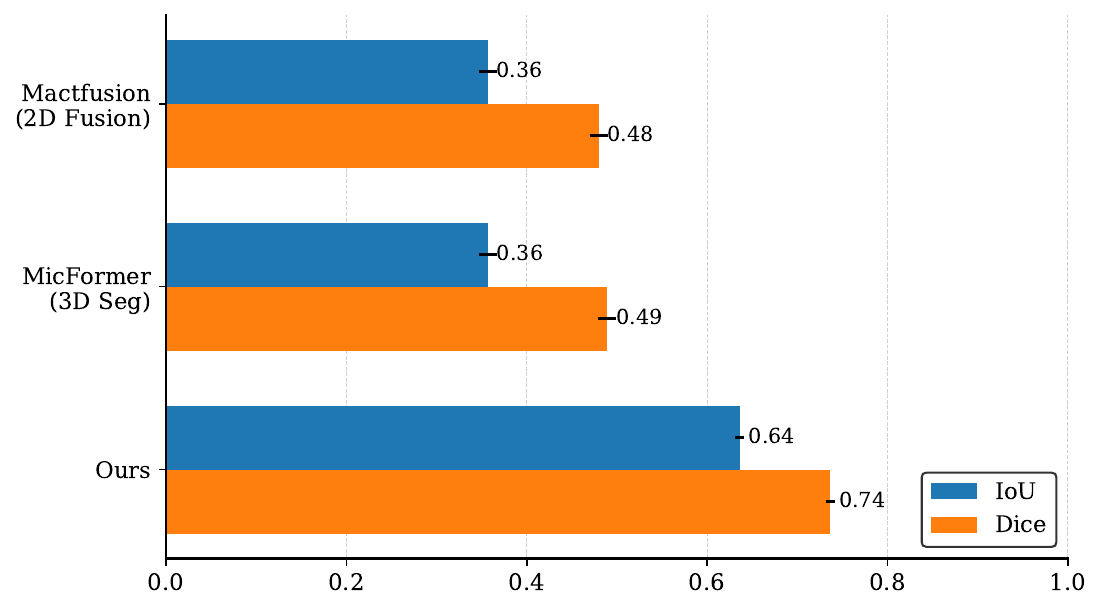}
    \caption{Performance of different methods (2D fusion method, 3D segmentation method, and ours) for multimodal 2D medical image segmentation tasks.}
    \label{fig:3d}
\end{figure}

Multimodal learning, which integrates complementary information from multiple modalities, has emerged as a promising solution to these challenges. By combining the macrostructural context of WLI with the microvascular precision of NBI, multimodal approaches aim to achieve a more comprehensive understanding of tumor characteristics. However, the integration of multimodal data introduces several technical challenges, including significant differences in image distributions, resolutions, and acquisition conditions across modalities~\citep{xiao2025decouple}. Effective fusion requires the alignment of modality-specific features while preserving the unique contributions of each modality, a task that is far from trivial.
Recently, studies on multimodal medical image segmentation have predominantly focused on 3D images, such as MRI, CT, and PET~\citep{huang2025evidentialfusion, yang2023learning, meng2024multi, zhang2022mmformer}. Typically, adapting segmentation methods developed for 3D multimodal images to 2D scenarios requires artificially augmenting 2D images into pseudo-3D formats. However, directly transferring these 3D fusion modules to forcibly model pseudo depth relationships may fail to capture 2D specific modal associations, misleading the model to learn invalid features. 
In addition, several methods have explored multimodal fusion specifically for 2D images~\citep{tang2022matr, wei2024ecinfusion, xie2024mactfusion, ma2022swinfusion, wang2025wavefusion}, however, these techniques are generally designed for 
pixel-level fusion rather than segmentation, which is often based on linear operations (such as weighted averaging), making it difficult to model complex pathological associations between modalities. In addition, the fusion process is usually independent of the segmentation task design and is not optimized for the segmentation target, resulting in the fusion image containing redundant information or lacking key discriminative features.
As shown in Fig.\ref{fig:3d}, for the 2D medical image segmentation task, the performance of the 3D method MicFormer~\citep{fan2024multimodal} and the 2D fusion technique Mactfusion~\citep{xie2024mactfusion} is not satisfactory compared to our proposed method specifically for multimodal 2D medical image segmentation. Additionally, some approaches adopt simple fusion strategies, such as concatenation before or after feature encoding, while others employ more sophisticated multi-scale fusion mechanisms. Nevertheless, these methods typically struggle to capture high-order cross-modal relationships. 

In this study, we present a novel 2D multimodal medical image segmentation framework tailored for laryngo-pharyngeal tumor recognition. This framework harnesses the complementary strengths of WLI and NBI through a multi-scale distribution alignment strategy, which aligns feature representations across modalities during the early encoding stage. Furthermore, a progressive disentanglement-aware multimodal contrastive learning strategy is presented, which effectively gradually separates modality-shared and modality-specific features on the basis of distribution alignment. This disentanglement approach enables fine-grained effective semantic fusion for segmentaion target due to separate processing of multimodal complementary shared features and modal private ones, ensuring that the integrated features are not only robust but also clinically effectiveness. 
The main contributions of this work are summarized as follows:

\begin{itemize}
    \item We propose a novel multimodal learning framework that effectively addresses cross-modal discrepancies by Align-Disentangle-Fusion mechanism for 2D medical image segmentation. By employing multi-scale distribution alignment on low-level features, the framework mitigates statistical mismatches. High-level features are then progressively disentangled into modality-shared and modality-specific components, enabling precise feature fusion to form a unified representation.
    \item We develop a disentangle-aware contrastive learning strategy based on the proposed framework, which simultaneously enhances the alignment of shared features across modalities and strengthens the separation between shared and specific features, facilitating robust multimodal image representation learning.
    \item Extensive experiments on three datasets are conducted to evaluate the proposed methods. The results consistently demonstrate significant improvements over other state-of-the-art approaches.
\end{itemize}


\section{Related Work}

\subsection{Medical Image Segmentation}

Medical image segmentation is a cornerstone of computer-aided diagnosis and has witnessed substantial advancements with the rise of deep learning. Convolutional neural networks (CNNs) have shown remarkable success in capturing local spatial features, with widely adopted architectures such as Unet~\citep{ronneberger2015u}. Numerous variants have been proposed to further improve segmentation performance, including~\citep{li2024dmsa,zhou2018unet++}. In addition to architectural variants, recent efforts have sought to enhance CNNs themselves~\citep{zhou2024uncertainty,qiao2024effective}. For instance, ConvFormer~\citep{lin2023convformer} adopts a Transformer-like architecture but employs convolution as the primary forward operator. Similarly, Shu et al.~\citep{shu2024csca} develope a channel and space compound attention mechanism to boost CNN expressiveness.

Recently, Transformer-based architectures have gained traction due to their ability to model long-range dependencies and capture global context, with notable contributions such as~\citep{he2023h2former, shaker2024unetrpp, wu2024medsegdiffv2}. For example, Unetr~\citep{hatamizadeh2021unetr} employs a pure Transformer encoder to process 3D volumetric data, effectively preserving spatial information and enhancing global semantic representation. Medsegdiffv2~\citep{wu2024medsegdiffv2} introduces a diffusion-based Transformer framework specifically designed for medical image segmentation. To harness the complementary advantages of CNNs and Transformers, a series of hybrid architectures have emerged that embed Transformer modules into convolutional backbones. These models typically utilize CNNs to extract fine-grained local features and Transformers to model global dependencies, enabling more expressive and robust representations for complex segmentation tasks~\citep{chen2021transunet, wu2024cnn, li2024scribformer, guo2024uctnet}. 

In head and neck tumor segmentation, URC~\citep{shi2023uncertainty} proposes a semi-supervised learning framework leveraging uncertainty estimation, while HECKTOR~\citep{andrearczyk2023automatic} introduces methods based on CT or fused PET/CT imaging. For laryngo-pharyngeal tumors, a subset of head and neck cancers, endoscopic imaging modalities such as WLI and NBI provide valuable diagnostic information. However, most existing studies focus solely on single-modality imaging~\citep{li2018nasopharyngeal,yumii2024laryngopharyngeal,de2024probability}, without explicitly employing complementary visual cues to enhance the recognition performance. 


\subsection{Multimodal Medical Image Analysis}

Early methods for 3D multimodal medical image segmentation primarily relied on simple fusion strategies, such as concatenating features from multiple modalities as input~\citep{menze2014multimodal}, but often failed to fully exploit cross-modal complementarities. To address these limitations, more advanced approaches have been proposed, Liu~\citep{liu2025completed} presents a dynamic Mixture-of-Experts Fusion module for explicitly learning the local-global feature relationships. mmFormer~\citep{zhang2022mmformer} leverages Transformer-based fusion to handle missing modalities. More recently, evidential fusion~\citep{huang2025evidentialfusion} introduced a DST-based framework that integrates uncertainty quantification and reliability modeling. However, directly adapting these multimodal 3D methods to 2D scenarios is suboptimal. Expanding a single 2D image into pseudo-3D input (such as repeated stacking or filling depth dimensions) may introduce false depth correlations, misleading the model to learn invalid features. Meantime, introducing unnecessary depth dimension calculations (such as pseudo-3D convolution) leads to a significant increase in computational and parameter complexity.

In addition to segmentation-focused frameworks, some methods have been proposed for multimodal 2D medical image fusion, aiming to generate a single more informative image by integrating complementary information across modalities. For instance, SwinFusion~\citep{ma2022swinfusion} unifies CNN and Swin Transformer architectures to model both intra- and inter-domain long-range dependencies. GeSeNet~\citep{li2024gesenet} presents a semantic-guided network that enhances fusion quality and efficiency for multimodal medical images. MACTFusion~\citep{xie2024mactfusion} introduces a lightweight cross-transformer framework for adaptive fusion, integrating local and global features across modalities. However, pixel-level image fusion generally requires strict modal alignment requirements, and the fusion process may not be able to preserve key features of all modalities. Additionally, the fusion process is usually independent of the segmentation task design and does not optimize the feature preservation strategy for the segmentation target, resulting in the fusion image containing redundant information or lacking key discriminative features. 

\subsection{Disentangled Representation Learning in Medical Image Analysis}

Recently, disentangled representation learning has been applied to various medical image analysis tasks as an effective strategy for learning interpretable features from original highly entangled feature space~\citep{zhou2024multi,liu2022disentangled,wang2022disentangled,liu2022learning}. Xiao et al.~\citep{xiao2025decouple} learns a decoupled anatomical representation for various tumor areas, which enhances the segmentation performances. Che et al.~\citep{che2025disentangle} present the DTF-Net which isolates multimodal features into low-frequency and high-frequency features, reducing the interference among them. Zhang et al. \citep{zhang2020exploring} propose a task-oriented segmentation framework with multi-modality feature representation learning, where attention-aware embedding operations generate context-weighted feature maps. Pei et al.~\citep{pei2021disentangle} learn disentangled domain feature for
domain adaptation and segmentation. Hu et al.~\citep{hu2020disentangled} propose a disentangled multimodal adversarial autoencoder that disentangle brain MRIs information into shared and complementary information by a VAE. Most existing assume strict alignment of multimodal data, so direct decoupling results are not optimal for data with significant modal differences, misalignment, or partial alignment. Different from most previous methods, in this work, we designed a decoupled learning strategy that follows a hierarchical refinement and sequential progression from distribution alignment, preliminary disentanglement, and disentangle-aware contrastive learning, which gradually improves the effectiveness of decoupling representation.


\section{Method}

\begin{figure*}[t]
    \centering
    \includegraphics[width=1\textwidth]{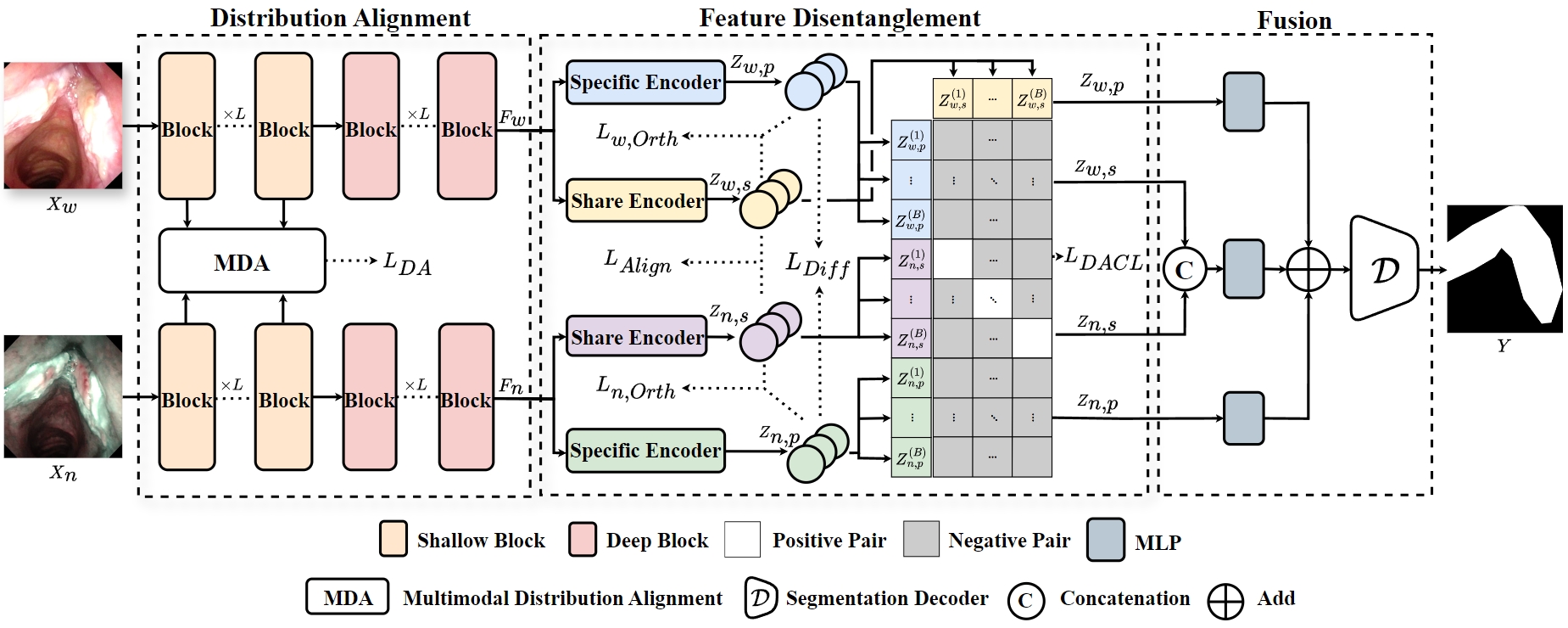} 
    \caption{Overview of the proposed framework. The distribution alignment is first conducted to bridge the feature distribution and semantic discrepancies and improve cross-modal compatibility. Second, the multimodal feature disentanglement is designed for explicitly modeling both complementary and modality-unique representations to enable more effective feature fusion.}
    \label{fig:1}
\end{figure*}

The overall pipeline is illustrated in Fig.~\ref{fig:1}. To effectively integrate the complementary information of WLI and NBI images, based on Alignment–Disentangle–Fusion strategy, we have designed and proposed an novel encoder-decoder learning framework for multimodal medical endoscopic image segmentation. The details are described as follows.

\subsection{Multimodal Distribution Alignment}
Although WLI and NBI offer complementary diagnostic insights, they also exhibit significant discrepancies such as spectral properties, illumination, and acquisition settings. These variations lead to distribution mismatches and modality-specific biases, hindering effective joint learning. To bridge this gap and improve cross-modal compatibility, we first perform feature-level distribution alignment.

In deep networks, shallow features primarily capture statistical distribution properties, while deeper layers encode high-level semantics. Therefore, we align distributions at the shallow encoder stages, where statistical disparities are more pronounced. 
In practice, we adopt a multi-scale approach that integrates global contextual cues with fine-grained local details, ensuring more robust cross-modal distribution alignment.

\subsubsection{Multi-Scale Feature Extraction}
To address these challenges, given a WLI-NBI pair $\{X_w,X_n\}$, we perform feature extraction at multiple shallow encoder stages and aggregate the extracted features into multi-scale representations for each modality. Let
\(\{\mathbf{f}_{w}^{\ell}\}_{\ell=1}^{L}\)
and
\(\{\mathbf{f}_{n}^{\ell}\}_{\ell=1}^{L}\)
denote the feature maps at \(L\) shallow stages for the WLI and NBI encoders, respectively. Each feature map has the form 
\(\mathbf{f}_{w}^{\ell}, \mathbf{f}_{n}^{\ell}\in\mathbb{R}^{B\times N\times D}\),
where \(B\) is the batch size, \(N\) denotes the number of spatial patches (or tokens), and \(D\) indicates the embedding dimension per patch.

To construct multi-scale representations, we concatenate the shallow-stage features along the embedding dimension:


\begin{equation}\label{eq:multi_scale_concat_app_revised}
\begin{aligned}
\mathbf{f}_{w} &= [\mathbf{f}_{w}^{1}; \mathbf{f}_{w}^{2}; \ldots; \mathbf{f}_{w}^{L}], \\
\mathbf{f}_{n} &= [\mathbf{f}_{n}^{1}; \mathbf{f}_{n}^{2}; \ldots; \mathbf{f}_{n}^{L}],
\end{aligned}
\end{equation}
\noindent where $[\,;\,]$ denotes concatenation along the last dimension, yielding $\mathbf{f}_{w}, \mathbf{f}_{n} \in \mathbb{R}^{B\times N\times (L \times D)}$. Because these features span coarse-to-fine semantic cues, they capture both global context and local details across modalities.

\subsubsection{Global Feature Extraction}

To facilitate robust and stable distribution alignment across modalities, we aim to construct a global feature representation that encapsulates both the statistical characteristics and the visual saliency. 
Guided by this objective, we define two complementary components within the global representation: one that captures the holistic statistical distribution of the input features, and another that selectively emphasizes spatial regions of semantic importance. 

To model the statistical aspect, we apply global average pooling over spatial tokens from the multi-scale feature maps $\mathbf{f}_{w}$ and $\mathbf{f}_{n}$, yielding compact descriptors that reflect the overall feature activation:
\begin{equation}\label{eq:avg_new_app_revised}
\mathbf{f}_{w}^{\mathrm{avg}} = \frac{1}{N} \sum_{n=1}^{N} \mathbf{f}_{w}[:, n, :], \quad
\mathbf{f}_{n}^{\mathrm{avg}} = \frac{1}{N} \sum_{n=1}^{N} \mathbf{f}_{n}[:, n, :],
\end{equation}
where $\mathbf{f}_{w}^{\mathrm{avg}}, \mathbf{f}_{n}^{\mathrm{avg}} \in \mathbb{R}^{B \times (L \times D)}$ serve as global descriptors capturing the statistical distribution of WLI and NBI features, respectively.


In parallel, to model the spatial and semantic relevance within the global context, we construct a complementary representation by introducing a weighted aggregation mechanism that emphasizes informative regions. Specifically, the multi-scale concatenated features $\mathbf{f}_{w}, \mathbf{f}_{n}$ are passed through a patch-wise scoring function to produce raw importance scores:
\begin{equation}\label{eq:attn_score_generation}
\mathbf{A}_{fw} = f_a^{(w)}(\mathbf{f}_{w}), \quad \mathbf{A}_{fn} = f_a^{(n)}(\mathbf{f}_{n}),
\end{equation}
where \( f_a^{(w)}(\cdot) \) and \( f_a^{(n)}(\cdot) \) denote separate token-level linear projections applied to WLI and NBI features, respectively. The resulting scores $\mathbf{A}_{fw}, \mathbf{A}_{fn} \in \mathbb{R}^{B \times N}$ are then normalized using the softmax function to produce attention weights:
\begin{equation}\label{eq:weighted_scores_new_app_revised}
w^{b,n}_{fw} = \frac{\exp(\mathbf{A}^{b,n}_{fw})}{\sum_{r=1}^{N} \exp(\mathbf{A}^{b,r}_{fw})}, \quad
w^{b,n}_{fn} = \frac{\exp(\mathbf{A}^{b,n}_{fn})}{\sum_{r=1}^{N} \exp(\mathbf{A}^{b,r}_{fn})}.
\end{equation}
These weights guide the aggregation of spatial features by assigning higher importance to more informative patches, yielding attention-aware global representations:
\begin{equation}\label{eq:weighted_feature_new_app_revised}
\mathbf{f}_{w}^{\mathrm{weighted}} = \sum_{n=1}^{N} w^{b,n}_{fw} \,\mathbf{f}^{b,n}_{w}, \quad
\mathbf{f}_{n}^{\mathrm{weighted}} = \sum_{n=1}^{N} w^{b,n}_{fn} \,\mathbf{f}^{b,n}_{n}.
\end{equation}
The resulting descriptors $\mathbf{f}_{w}^{\mathrm{weighted}}, \mathbf{f}_{n}^{\mathrm{weighted}} \in \mathbb{R}^{B \times (L \times D)}$ emphasize semantically meaningful regions and contribute to more discriminative global representations.

Finally, we integrate the statistical and attention-weighted descriptors to form the complete global representations:
\begin{equation}\label{eq:s_global_new_app_revised}
\mathbf{f}_{w}^{\mathrm{Global}} = \mathbf{f}_{w}^{\mathrm{avg}} + \mathbf{f}_{w}^{\mathrm{weighted}}, \quad
\mathbf{f}_{n}^{\mathrm{Global}} = \mathbf{f}_{n}^{\mathrm{avg}} + \mathbf{f}_{n}^{\mathrm{weighted}}.
\end{equation}
This unified representation captures both global distributional structure and spatial saliency, offering a more expressive and semantically robust foundation for subsequent cross-modal distribution alignment.

\subsubsection{Distribution Alignment}
After obtaining the global representations, we apply Maximum Mean Discrepancy (MMD) to align feature distributions and mitigate modality discrepancies. MMD effectively measures distribution differences in high-dimensional spaces by mapping features to a reproducing kernel Hilbert space, ensuring both mean and higher-order statistical consistency. 
Given the final global representations 
\(\mathbf{f}_{w}^{\mathrm{Global}}, \mathbf{f}_{n}^{\mathrm{Global}} \in \mathbb{R}^{B\times (L \times D)}\), 
we apply MMD with Gaussian kernel to align the two distributions:
\begin{equation}\label{eq:mmd_loss_global_app_revised}
\begin{aligned}
\mathcal{L}_{\mathrm{DA}}
&= \frac{1}{B^2} \sum_{i=1}^{B} \sum_{j=1}^{B} 
   \exp\Bigl(-\tfrac{\|\mathbf{f}_{w,i}^{\mathrm{Global}} - \mathbf{f}_{w,j}^{\mathrm{Global}}\|^2}{2\sigma^2}\Bigr) \\
\quad &+ \frac{1}{B^2} \sum_{p=1}^{B} \sum_{q=1}^{B} 
   \exp\Bigl(-\tfrac{\|\mathbf{f}_{n,p}^{\mathrm{Global}} - \mathbf{f}_{n,q}^{\mathrm{Global}}\|^2}{2\sigma^2}\Bigr) \\
\quad &- \frac{2}{B^2} \sum_{i=1}^{B} \sum_{p=1}^{B} 
   \exp\Bigl(-\tfrac{\|\mathbf{f}_{w,i}^{\mathrm{Global}} - \mathbf{f}_{n,p}^{\mathrm{Global}}\|^2}{2\sigma^2}\Bigr),
\end{aligned}
\end{equation}
where \(\sigma\) is the kernel bandwidth. By minimizing \(\mathcal{L}_{\mathrm{DA}}\), we enforce feature alignment between WLI and NBI, which enhances cross-modal feature fusion and lead to more robust representations.


\subsection{Multimodal Feature Disentanglement}
\begin{figure}[h]
    \centering
    \includegraphics[width=0.3\textwidth]{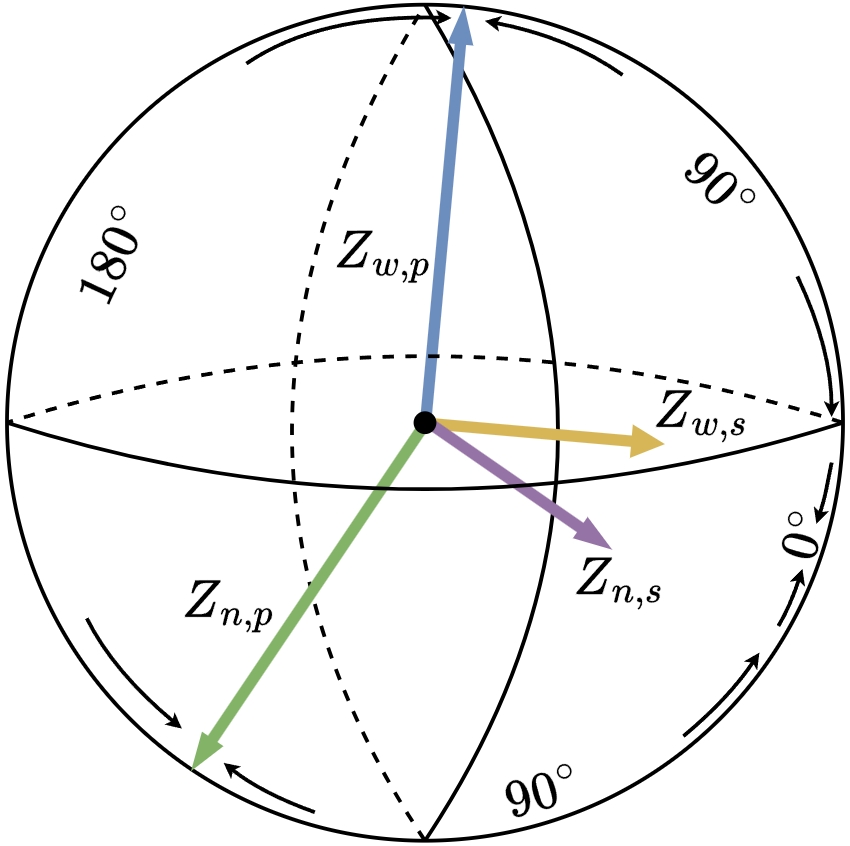}
    \caption{Schematic of the geometric properties of various feature similarity in preliminary disentanglement learning.}
    \label{fig:PD}
\end{figure}

Even after distribution alignment, semantic discrepancies between modalities may still persist due to the inherent differences in modality-specific information. These discrepancies are often reflected in redundant and complementary semantic features across modalities. To better leverage these semantic cues, we further decouple the aligned features into a shared subspace and modality-specific subspaces, explicitly modeling both complementary and modality-unique representations to enable more effective feature fusion.

\subsubsection{Feature Extraction}
Let \(\mathbf{F}_{\mathrm{w}}\) and \(\mathbf{F}_{\mathrm{n}}\) denote the last encoded features from the WLI and NBI modalities, respectively. To capture both shared and specific features of each modality, we employ different encoders $\{f_\mathrm{s}^{(i)}$, $f_\mathrm{p}^{(i)}\}$ to extract modality-shared and modality-specific features of modality $i$. Formally, we define:
\begin{equation}
\begin{aligned}
\mathbf{Z}_{w,s} &= f_{s}^{(w)}(\mathbf{F}_{\mathrm{w}}), 
\quad 
\mathbf{Z}_{w,\mathrm{p}} = f_{\mathrm{p}}^{(w)}(\mathbf{F}_{\mathrm{w}}),\\
\mathbf{Z}_{n,s} &= f_{s}^{(n)}(\mathbf{F}_{\mathrm{n}}), 
\quad 
\mathbf{Z}_{n,\mathrm{p}} = f_{\mathrm{p}}^{(n)}(\mathbf{F}_{\mathrm{n}}),
\end{aligned}
\end{equation}
where \(\mathbf{Z}_{i,s}\) denotes the shared feature component of modality \(i\), and \(\mathbf{Z}_{i,\mathrm{p}}\) denotes its modality-specific feature component.

\subsubsection{Preliminary Disentanglement}
We explicitly model the relationships between modality-shared and modality-specific features. Specifically, we assume that the shared features across different modalities should be well-aligned, the shared and specific features within the same modality should remain mutually independent, and the modality-specific features from different modalities should also be independent with weak correlations. To enforce these relational constraints, we introduce the geometric properties of different feature similarity as a measure of semantic relationships between feature vectors in the embedding space, which is shown in Fig. \ref{fig:PD}. Specifically, cosine similarity characterizes the angular relationship between feature vectors, where different angles correspond to varying degrees of semantic correlation. Based on this property, we design a set of constraints to regulate the similarity between features, thereby achieving preliminary feature decoupling and laying the foundation for subsequent feature fusion.

\paragraph{Shared Feature Alignment}
To promote semantic consistency across modalities, we impose a constraint that encourages similarity between the shared features of WLI and NBI. In terms of cosine similarity, this corresponds to minimizing the angle between shared feature vectors, ideally approaching \(0^\circ\), which indicates maximal alignment. The corresponding constraint is formulated as:
\begin{equation}
\mathcal{L}_{\mathrm{Align}} 
= \frac{1}{2} \Biggl(1 - \frac{1}{B}\sum_{b=1}^B
\frac{\mathbf{Z}_{w,s}^{(b)} \cdot \mathbf{Z}_{n,s}^{(b)}}%
{\|\mathbf{Z}_{w,s}^{(b)}\|\,\|\mathbf{Z}_{n,s}^{(b)}\|}\Biggr).
\end{equation}

\paragraph{Cross-Modal Specific Feature Differentiation}
To maximize the distinction between modality-specific features of WLI and NBI, we impose a constraint that minimizes their similarity, encouraging their directions to approach \(180^\circ\) in the embedding space. In this case, the cosine similarity tends towards \(-1\), indicating maximal semantic opposition. The corresponding constraint is defined as:
\begin{equation}
\mathcal{L}_{\mathrm{Diff}}
= \frac{1}{2}\Bigl(
1 + \frac{1}{B}\sum_{b=1}^B
\frac{\mathbf{Z}_{w,\mathrm{p}}^{(b)} \cdot \mathbf{Z}_{n,\mathrm{p}}^{(b)}}%
{\|\mathbf{Z}_{w,\mathrm{p}}^{(b)}\|\|\mathbf{Z}_{n,\mathrm{p}}^{(b)}\|}
\Bigr).
\end{equation}

\paragraph{Intra-Modal Orthogonality}
To disentangle shared and specific features within each modality, we impose an orthogonality constraint between the shared and modality-specific components of WLI and NBI. Geometrically, this corresponds to a \(90^\circ\) angle between the respective feature vectors, ensuring that shared and specific features capture distinct semantic subspaces. The corresponding constraint is expressed as:
\begin{equation}
\mathcal{L}_{\mathrm{Orth}} 
= \frac{1}{2B}\sum_{b=1}^B
\Bigl[
\underbrace{\Bigl(\frac{\mathbf{Z}_{w,s}^{(b)} \cdot \mathbf{Z}_{w,\mathrm{p}}^{(b)}}%
{\|\mathbf{Z}_{w,s}^{(b)}\|\|\mathbf{Z}_{w,\mathrm{p}}^{(b)}\|}
\Bigr)^2}_{\textbf{$\mathcal{L}_{w,\mathrm{Orth}}$}}
+ 
\underbrace{\Bigl(\frac{\mathbf{Z}_{n,s}^{(b)} \cdot \mathbf{Z}_{n,\mathrm{p}}^{(b)}}%
{\|\mathbf{Z}_{n,s}^{(b)}\|\|\mathbf{Z}_{n,\mathrm{p}}^{(b)}\|}
\Bigr)^2}_{\textbf{$\mathcal{L}_{n,\mathrm{Orth}}$}}
\Bigr].
\end{equation}

\subsubsection{Disentangle-aware Contrastive Learning}

To further strengthen the discriminability of disentangled representations, we introduce a contrastive learning strategy based on multimodal disentangled features. After the initial decoupling, we obtain the feature set 
\(\{\mathbf{Z}_{w,s}, \mathbf{Z}_{w,\mathrm{p}}, \\\mathbf{Z}_{n,s}, \mathbf{Z}_{n,\mathrm{p}}\}\). In this framework, for each sample \(b\), we explicitly pull together the shared features from different modalities (\(\mathbf{Z}_{w,s}^{(b)}\) and \(\mathbf{Z}_{n,s}^{(b)}\)) by treating them as positive pairs, as they are expected to capture common semantic information. Simultaneously, we push apart the shared features and the modality-specific features from both modalities, treating modality-specific features (\(\mathbf{Z}_{w,\mathrm{p}}^{(b)}\) and \(\mathbf{Z}_{n,\mathrm{p}}^{(b)}\)) as negative samples. This design promotes compactness within the shared feature space while increasing the separation between shared and specific representations, thereby facilitating more thorough feature disentanglement. To achieve this, we propose Disentangle-aware Contrastive Learning (DACL), which encourages the alignment of positive pairs (\emph{i.e.}, shared features across modalities) while effectively pushing apart multiple negative samples, including modality-specific features and irrelevant shared features.
\begin{equation}
\mathcal{L}_{\mathrm{DACL}} 
= -\frac{1}{B} \sum_{b=1}^B 
\log \Biggl(\frac{\mathcal{S}_{\mathrm{pos}}^{(b)}}{\mathcal{S}_{\mathrm{den}}^{(b)}}\Biggr),
\end{equation}
where the positive similarity score is defined as:
\begin{equation}
\mathcal{S}_{\mathrm{pos}}^{(b)} 
= \exp\Biggl(
    \frac{
        \mathrm{sim}(\mathbf{Z}_{w,s}^{(b)}, \mathbf{Z}_{n,s}^{(b)})
    }{\tau}
\Biggr),
\end{equation}
and the denominator aggregates all similarities as:
\begin{equation}
\begin{aligned}
\mathcal{S}_{\mathrm{den}}^{(b)} 
&= \sum_{m=1}^B \Biggl[
  \exp\Biggl(
      \frac{
          \mathrm{sim}(\mathbf{Z}_{w,s}^{(b)}, \mathbf{Z}_{n,s}^{(m)})
      }{\tau}
  \Biggr) \\
&\quad + \exp\Biggl(
      \frac{
          \mathrm{sim}(\mathbf{Z}_{w,s}^{(b)}, \mathbf{Z}_{w,\mathrm{p}}^{(m)})
      }{\tau}
  \Biggr) \\
&\quad + \exp\Biggl(
      \frac{
          \mathrm{sim}(\mathbf{Z}_{w,s}^{(b)}, \mathbf{Z}_{n,\mathrm{p}}^{(m)})
      }{\tau}
  \Biggr)
\Biggr],
\end{aligned}
\end{equation}
where \(\mathrm{sim}(\mathbf{a}, \mathbf{b}) = \frac{\mathbf{a} \cdot \mathbf{b}}{\|\mathbf{a}\| \|\mathbf{b}\|}\) denotes the cosine similarity, and \(\tau>0\) is a temperature parameter controlling the concentration level of the distribution. By minimizing \(\mathcal{L}_{\mathrm{DACL}}\), the shared features from the same sample across modalities are drawn closer, while their distances to modality-specific features and unrelated shared features are increased, promoting stronger disentanglement.

\subsubsection{Overall Objective}
Combining all constraints and contrastive learning, the overall training objective is defined as:
\begin{equation}
\begin{array}{c}
\mathcal{L}_{\mathrm{FD}}
= \alpha\,\mathcal{L}_{\mathrm{Align}}
+ \beta\,\mathcal{L}_{\mathrm{Diff}}
+ \gamma\,\mathcal{L}_{\mathrm{Orth}}
+ \delta\,\mathcal{L}_{\mathrm{DACL}}, \\[8pt]
\alpha, \beta, \gamma, \delta \geq 0.
\end{array}
\end{equation}
To ensure numerical stability and prevent disproportionate loss scaling, in practice, we define the weighting parameters as \(\alpha = \beta = \gamma = \frac{1}{3}\) and \(\delta = 0.01\) (see sec. IV-E for the detailed hyperparameter analysis).

\subsection{Multimodal Feature Fusion}  
To fully exploit the disentangled multimodal features and enhance segmentation performance, we design the fusion process in two stages.  

\subsubsection{Cross-Modal Shared Feature Aggregation}  
After obtaining the disentangled features, we first enhance the shared feature representation by aligning the shared features from different modalities. Specifically, the shared features from WLI and NBI are denoted as \(\mathbf{Z}_{w,s}, \mathbf{Z}_{n,s} \in \mathbb{R}^{B \times N \times D}\), respectively. These features are concatenated along the feature dimension and projected into a unified subspace via a linear mapping:
\begin{equation}\label{eq:shared_feature_alignment}
    \mathbf{Z}_{\mathrm{sh}} = f_{\mathrm{sh}}\bigl( [\mathbf{Z}_{w,s}; \mathbf{Z}_{n,s}] \bigr),
\end{equation}
where \( [\cdot;\cdot] \) denotes concatenation along the last dimension, resulting in a tensor of shape \( \mathbb{R}^{B \times N \times 2D} \), and \( f_{\mathrm{sh}}: \mathbb{R}^{B \times N \times 2D} \to \mathbb{R}^{B \times N \times D} \) denotes a linear transformation applied independently to each patch.

\subsubsection{Fusion of Shared and Modality-Specific Features}  
The shared feature \(\mathbf{Z}_{\mathrm{sh}} \in \mathbb{R}^{B \times N \times D}\), together with the modality-specific features \(\mathbf{Z}_{w,\mathrm{p}}, \mathbf{Z}_{n,\mathrm{p}} \in \mathbb{R}^{B \times N \times D}\), are integrated to form the fused feature through the following operation:
\begin{equation}\label{eq:feature_fusion}
    \mathbf{Z}_{\mathrm{fused}} = f_{\mathrm{sh}}^{\prime}(\mathbf{Z}_{\mathrm{sh}}) + f_{w}(\mathbf{Z}_{w,\mathrm{p}}) + f_{n}(\mathbf{Z}_{n,\mathrm{p}}),
\end{equation}
where \( f_{\mathrm{sh}}^{\prime}, f_{w}, f_{n}: \mathbb{R}^{B \times N \times D} \to \mathbb{R}^{B \times N \times D} \) are independent non-linear transformations applied over the feature dimension of each patch. The resulting fused feature \( \mathbf{Z}_{\mathrm{fused}} \in \mathbb{R}^{B \times N \times D} \) aggregates complementary information from both modalities.

\subsubsection{Lesion Segmentation Prediction}  
Finally, the fused feature \(\mathbf{Z}_{\mathrm{fused}} \in \mathbb{R}^{B \times N \times D}\) is passed into the segmentation decoder \( \mathcal{D}: \mathbb{R}^{B \times N \times D} \to \mathbb{R}^{B \times H \times W} \) to predict the lesion mask \( Y \):
\begin{equation}\label{eq:segmentation_prediction}
    Y = \mathcal{D}(\mathbf{Z}_{\mathrm{fused}}).
\end{equation}
This fusion strategy effectively integrates both modality-shared and modality-specific representations, enabling accurate and robust lesion segmentation.

\subsection{Optimization}
\subsubsection{Overall Objective}
We integrate multimodal distribution alignment, feature disentanglement, and segmentation supervision into a unified overall training objective. 
Formally, the total loss is defined as:
\begin{equation}
\mathcal{L}_{\mathrm{total}} 
= \lambda_1\,\mathcal{L}_{\mathrm{DA}}
+ \lambda_2\,\mathcal{L}_{\mathrm{FD}}
+ \lambda_3\,\mathcal{L}_{\mathrm{CE}}
+ \lambda_4\,\mathcal{L}_{\mathrm{Dice}},
\end{equation}
where \(\{\lambda_1, \lambda_2, \lambda_3, \lambda_4\}\) are non-negative hyperparameters controlling the contribution of each component. Specifically:
\begin{itemize}
    \item \(\mathcal{L}_{\mathrm{DA}}\) aligns the distribution information between WLI and NBI, reducing modality discrepancies and ensuring consistent statistical properties across modalities in the shallow layers.
    \item \(\mathcal{L}_{\mathrm{FD}}\) A enforces disentanglement constraints, including shared feature alignment, modality-specific feature differentiation, and intra-modal orthogonality between the shared- and specific ones.
    \item \(\mathcal{L}_{\mathrm{CE}}\) and \(\mathcal{L}_{\mathrm{Dice}}\) provide pixel-level supervision to guide accurate lesion segmentation.
\end{itemize}
Minimizing \(\mathcal{L}_{\mathrm{total}}\) facilitates effective cross-modal feature alignment and disentanglement, while ensuring robust and precise segmentation performance.

\subsubsection{Progressive Loss Weighting}
\begin{algorithm}[H]
\caption{Progressive Loss Weighting Procedure}
\label{alg:progressive_loss}
\begin{algorithmic}[1]
\State $E \gets \text{total number of epochs}$
\State $\lambda_1 \gets \alpha_{\mathrm{DA}},\; \lambda_3 \gets \alpha_{\mathrm{CE}},\; \lambda_4 \gets \alpha_{\mathrm{Dice}}$ \hfill \textbf{// Fixed weights for DA, CE, and Dice}
\State $(\alpha_{\mathrm{FD},\max}, \alpha_{\mathrm{FD},\mathrm{init}})$ 
\hfill \textbf{// For dynamic adjustment of \(\lambda_2\)}
\For{$e = 1$ \textbf{to} $E$}
    \State $\lambda_2 \gets \min\Bigl(\alpha_{\mathrm{FD},\max}, \frac{e}{E} \cdot \alpha_{\mathrm{FD},\mathrm{init}}\Bigr)$
    \Comment{Progressive weight for \(\mathcal{L}_{\mathrm{FD}}\)}
    \State \textbf{Compute} $\mathcal{L}_{\mathrm{DA}}, \mathcal{L}_{\mathrm{CE}}, \mathcal{L}_{\mathrm{Dice}}$
    \State $\mathcal{L}_{\mathrm{FD}} \gets 
        \alpha\,\mathcal{L}_{\mathrm{Align}}
        + \beta\,\mathcal{L}_{\mathrm{Diff}}
        + \gamma\,\mathcal{L}_{\mathrm{Orth}}
        + \delta\,\mathcal{L}_{\mathrm{DACL}}$
    \Comment{Fixed internal weights \(\alpha, \beta, \gamma, \delta\)}
    \State $\mathcal{L}_{\mathrm{total}} \gets 
        \lambda_{1}\mathcal{L}_{\mathrm{DA}}
        + \lambda_{2}\mathcal{L}_{\mathrm{FD}}
        + \lambda_{3}\mathcal{L}_{\mathrm{CE}}
        + \lambda_{4}\mathcal{L}_{\mathrm{Dice}}$
    \State \textbf{Update model parameters} by minimizing $\mathcal{L}_{\mathrm{total}}$
\EndFor
\end{algorithmic}
\end{algorithm}

Directly optimizing multiple objectives with distinct goals may introduce conflicts between loss terms, leading to suboptimal performance. To mitigate this, we design a progressive weighting scheme that gradually balances the contributions of each loss during training. In the early stages, we prioritize segmentation supervision (\(\mathcal{L}_{\mathrm{CE}}, \mathcal{L}_{\mathrm{Dice}}\)) and distribution alignment (\(\mathcal{L}_{\mathrm{DA}}\)) to ensure stable feature representations and reliable cross-modal consistency. Once the model establishes a solid foundation in segmentation and distribution alignment, we progressively emphasize feature decomposition (\(\mathcal{L}_{\mathrm{FD}}\)) to disentangle high-level semantic features and enhance multimodal fusion. This staged optimization strategy allows the model to first focus on fundamental tasks before refining its multimodal understanding, as detailed in Algorithm~\ref{alg:progressive_loss}.


\section{Experienments}

\subsection{Dataset}

In this study, as shown in Table \ref{tab:dataset_distribution}, experiments are conducted on three clinical multimodal endoscopic image datasets (\emph{i.e.}, \textbf{Dataset-I}, \textbf{Dataset-II}, and \textbf{Dataset-III}) with paired WLI and NBI images of the laryngo-pharyngeal. These three datasets were collected from the First Affiliated Hospital of Sun Yat-sen University, the Sun Yat-sen Memorial Hospital and the Sixth Affiliated Hospital of Sun Yat-sen University, Guangzhou, China. In our work, the WLI images is considered as the primary modality and the corresponding label is as the reference ground-truth for WLI-NBI pairs. All WLI images are annotated by clinicians. Specifically, \textbf{Dataset-I} consists of $2,209$ pairs of multi-modal images with $1,808$ pairs used for training and $401$ pairs for testing. In \textbf{Dataset-II}, there are $333$ pairs of WLI-NBI images, $266$ pairs for training and $67$ pairs for test. \textbf{Dataset-III} contains $81$ WLI-NBI pairs, divided into $64$ pairs for training and $17$ pairs for test. For fair comparisons, the patient IDs in all the training set are different from those in the test set in above three datasets. 
\begin{table}[htbp]
\centering
\caption{Sample distribution in the training and testing sets across the three datasets.}
\label{tab:dataset_distribution}
\resizebox{0.48\textwidth}{!}{%
\begin{tabular}{c|cc|cc|cc}
\toprule
\multicolumn{1}{c|}{\multirow{2}[3]{*}{\textbf{Category}}}
& \multicolumn{2}{c|}{\textbf{Dataset-I}} & \multicolumn{2}{c|}{\textbf{Dataset-II}} & \multicolumn{2}{c}{\textbf{Dataset-III}} \\
\cmidrule{2-7}
 & \textbf{Train} & \textbf{Test} & \textbf{Train} & \textbf{Test} & \textbf{Train} & \textbf{Test} \\
\midrule
Benign & 758  & 146 & 170 & 41  & 18 & 5 \\
Tumor  & 1,050 & 255 & 96  & 26  & 46 & 12 \\
\midrule
Total  & 1,808 & 401 & 266 & 67  & 64 & 17 \\
\bottomrule
\end{tabular}%
}
\end{table}

\begin{figure*}[htbp]
    \centering
    \includegraphics[width=0.85\textwidth]{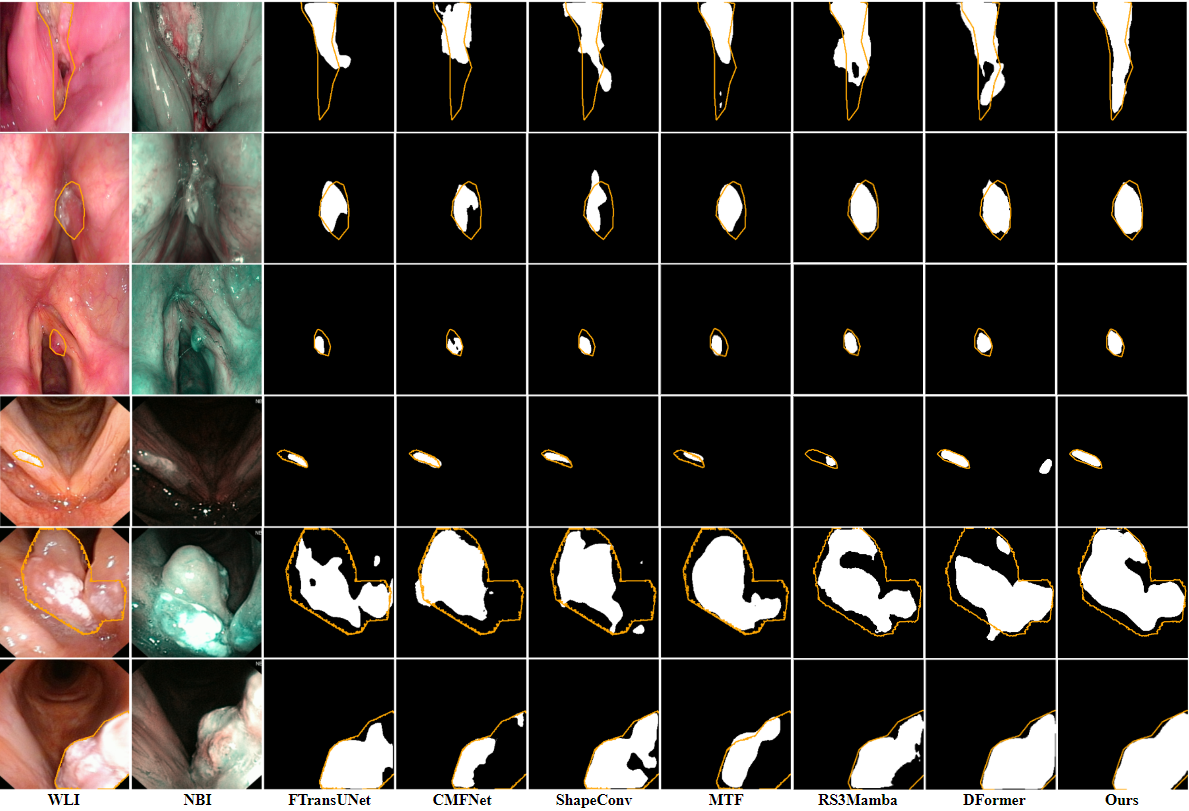}
    \captionsetup{justification=centering}
    \caption{Visualization comparisions with different approaches. The orange curve represents the ground-truth contour.}
    \label{fig:sota_visualization}
\end{figure*}

\subsection{Evaluation Metrics}

To comprehensively evaluate the segmentation performance, we adopt four widely used metrics: Intersection over Union (IoU), Dice coefficient (Dice), Sensitivity (SE), and Geometric Mean (G-mean)~\citep{he2009learning}. IoU and Dice assess the region-level overlap between predictions and ground truth, providing a general measure of segmentation accuracy. In this work, the IoU and Dice are as the primary indicators for evaluation.

\subsection{Training and Testing Details}
All experiments were conducted using PyTorch on one NVIDIA RTX 8000 GPU with 48GB of memory. In our work, we employ the TransUNet~\citep{chen2021transunet} encoder as the backbone. Input images were resized to $224 \times 224$ during both training and testing. The training process was carried out for $150$ epochs using the Adam optimizer, with a learning rate of \(1 \times 10^{-3}\) and a batch size of $24$. 
For the segmentation task, we assigned equal weights of 0.5 to the Dice loss and Cross-Entropy loss, emphasizing accurate segmentation as the primary objective. To align feature distributions across modalities, we set a small hyperparameter value of \(1 \times 10^{-4}\) for the distribution alignment loss, which proved effective without negatively impacting segmentation performance.

\begin{table}[htbp]
\centering
\caption{Quantitative comparison with state-of-the-art methods on \textbf{Dataset-I}.}
\label{tab:dataset1_comparison}
\resizebox{0.5\textwidth}{!}{%
\begin{tabular}{c|cc|cccc}
\toprule
\multicolumn{1}{c}{\multirow{2}[3]{*}{\textbf{Methods}}} & \multicolumn{2}{|c|}{\textbf{Modality}} & \multicolumn{4}{c}{\textbf{Metrics}} \\
\cmidrule{2-7}
& \textbf{WLI} & \textbf{NBI} & \textbf{IoU} & \textbf{Dice} & \textbf{SE} & \textbf{G-mean} \\
\midrule
TransUNet (2021)~\citep{chen2021transunet}          & \checkmark &            & 0.5852 & 0.6944 & 0.8188 & 0.8890 \\
TransUNet (2021)~\citep{chen2021transunet}          &            & \checkmark & 0.2713 & 0.3727 & 0.5738 & 0.7259 \\
\midrule
ShapeConv (2021)~\citep{ShapeConv}     & \checkmark & \checkmark & 0.6358 & 0.7435 & 0.7790 & 0.8699 \\
MTF (2022)~\citep{wang2022multimodaltokenfusionvision} & \checkmark & \checkmark & 0.6387 & 0.7459 & 0.7913 & 0.8771 \\
CMFNet (2022)~\citep{ma2022cmfnet}     & \checkmark & \checkmark & 0.5707 & 0.6829 & 0.6994 & 0.8254 \\
RS3Mamba (2024)~\citep{ma2024rs3mamba} & \checkmark & \checkmark & 0.6164 & 0.7284 & 0.7840 & 0.8772 \\
FTransUNet (2024)~\citep{ma2024fusion} & \checkmark & \checkmark & 0.6269 & 0.7355 & 0.7625 & 0.8604 \\
Dformer (2024)~\citep{yin2023dformer}  & \checkmark & \checkmark & 0.6313 & 0.7361 & 0.7657 & 0.8602 \\
\textbf{Ours}                         & \checkmark & \checkmark & \textbf{0.6453} & \textbf{0.7517} & \textbf{0.8560} & \textbf{0.9100} \\
\bottomrule
\end{tabular}}
\end{table}

\begin{table}[htbp]
\centering
\caption{Quantitative comparison with state-of-the-art methods on \textbf{Dataset-II}.}
\label{tab:dataset2_comparison}
\resizebox{0.5\textwidth}{!}{%
\begin{tabular}{c|cc|cccc}
\toprule
\multicolumn{1}{c}{\multirow{2}[3]{*}{\textbf{Methods}}} & \multicolumn{2}{|c|}{\textbf{Modality}} & \multicolumn{4}{c}{\textbf{Metrics}} \\
\cmidrule{2-7}
& \textbf{WLI} & \textbf{NBI} & \textbf{IoU} & \textbf{Dice} & \textbf{SE} & \textbf{G-mean} \\
\midrule
TransUNet (2021)~\citep{chen2021transunet}          & \checkmark &            & 0.7421 & 0.8283 & 0.8932 & 0.9424 \\
TransUNet (2021)~\citep{chen2021transunet}          &            & \checkmark & 0.4868 & 0.5674 & 0.7897 & 0.8769 \\
\midrule
ShapeConv (2021)~\citep{ShapeConv}     & \checkmark & \checkmark & 0.8007 & 0.8672 & 0.8825 & 0.9365 \\
MTF (2022)~\citep{wang2022multimodaltokenfusionvision} & \checkmark & \checkmark & 0.7526 & 0.8162 & 0.8354 & 0.9101 \\
CMFNet (2022)~\citep{ma2022cmfnet}     & \checkmark & \checkmark & 0.7940 & 0.8633 & 0.8885 & 0.9394 \\
RS3Mamba (2024)~\citep{ma2024rs3mamba} & \checkmark & \checkmark & 0.7133 & 0.8185 & 0.8462 & 0.9199 \\
FTransUNet (2024)~\citep{ma2024fusion} & \checkmark & \checkmark & 0.7839 & 0.8542 & 0.8581 & 0.9226 \\
Dformer (2024)~\citep{yin2023dformer}  & \checkmark & \checkmark & 0.7962 & 0.8615 & 0.9198 & 0.9562 \\
\textbf{Ours}                         & \checkmark & \checkmark & \textbf{0.8205} & \textbf{0.8895} & \textbf{0.9237} & \textbf{0.9599} \\
\bottomrule
\end{tabular}
}
\end{table}

\begin{figure}[htbp]
    \centering
    \includegraphics[width=0.5\textwidth]{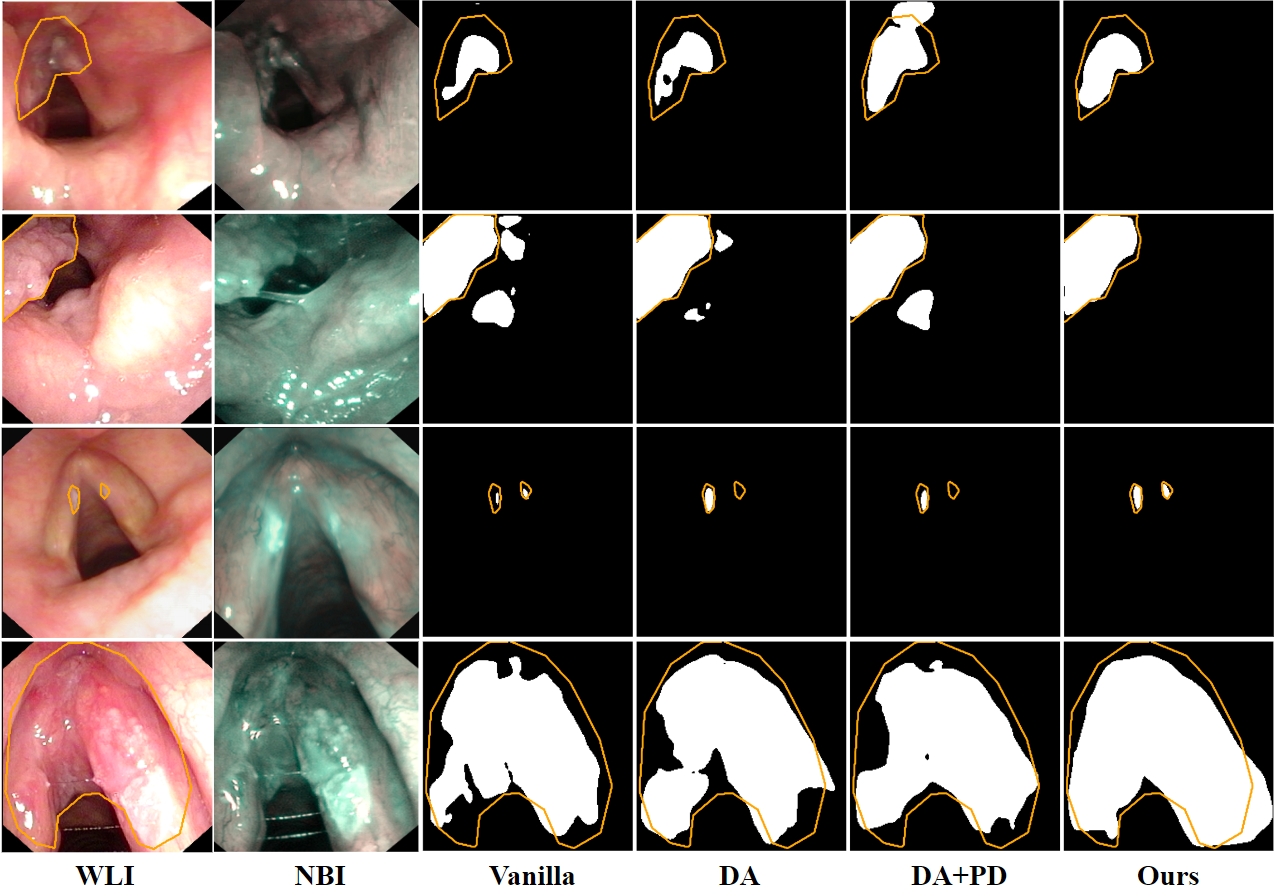}
    \caption{Ablation visualization. Orange curve represents the ground-truth contour.}
    \label{fig:5}
\end{figure}
\subsection{Comparison With State-of-the-Art Methods}

To assess the effectiveness of the proposed framework, extensive comparative experiments were systematically conducted against several state-of-the-art 2D multimodal image segmentation approaches, including ShapeConv~\citep{ShapeConv}, MTF~\citep{wang2022multimodaltokenfusionvision}, CMFNet~\citep{ma2022cmfnet}, RS3Mamba~\citep{ma2024rs3mamba}, FTransUNet~\citep{ma2024fusion}, and Dformer~\citep{yin2023dformer}.

First, we conducted single-modality experiments on multiple datasets. Since the ground truth labels were annotated based on WLI images, partial misalignment exists between the NBI images and their corresponding labels, which adversely impacts the performance of NBI-based single-modality segmentation. Despite this limitation, our proposed multimodal approach consistently outperforms single-modality methods across all datasets. Specifically, our method achieves improvements of $6.01\%$, $7.84\%$, and $10.4\%$ on the three datasets compared to the results based on the WLI modal, respectively, demonstrating the effectiveness of the proposed multimodal 2D medical image segmentation method.

The comparative quantitative results, comprehensively presented in Tables~\ref{tab:dataset1_comparison}--\ref{tab:dataset3_comparison}, reveal a consistent and significant performance advantage of our approach across diverse experimental settings. On \textbf{Dataset-I}, which contains a large amount of training data, our model outperformed all state-of-the-art methods, achieving substantial improvements of $7.46\%$ and $6.88\%$ in IoU and Dice scores over the CMFNet. Compared to the best competitor MTF, our method exceeds it by $0.66\%$ and $0.58\%$ in the IoU and Dice scores, respectively. These results validate the effectiveness of our method in data-rich scenarios, underscoring its potential capacity for disentangling feature representations.

Notably, our framework exhibited even greater advantages in data-limited settings. On \textbf{Dataset-II}, the proposed method outperformed the previous best-performing ShapeConv with gains of $1.98\%$ in IoU and $2.23\%$ in Dice. Similarly, on \textbf{Dataset-III}, our approach surpassed DFormer—the most competitive baseline—by $3.84\%$ and $3.24\%$ in IoU and Dice, respectively. These findings demonstrate that our feature disentanglement-based framework is particularly advantageous under data-scarce conditions, enabling more effective capture of salient features and highlighting its efficacy and robustness.

Furthermore, qualitative comparisons corroborate the superior segmentation capability of our model. As shown in Figure~\ref{fig:sota_visualization}, our method is more accurate in locating and recognizing irregular edges, and also more precise in identifying small tumors, which obtains outstanding segmentation performance. 

\begin{figure*}[htbp]
    \centering
    \includegraphics[width=1\textwidth]{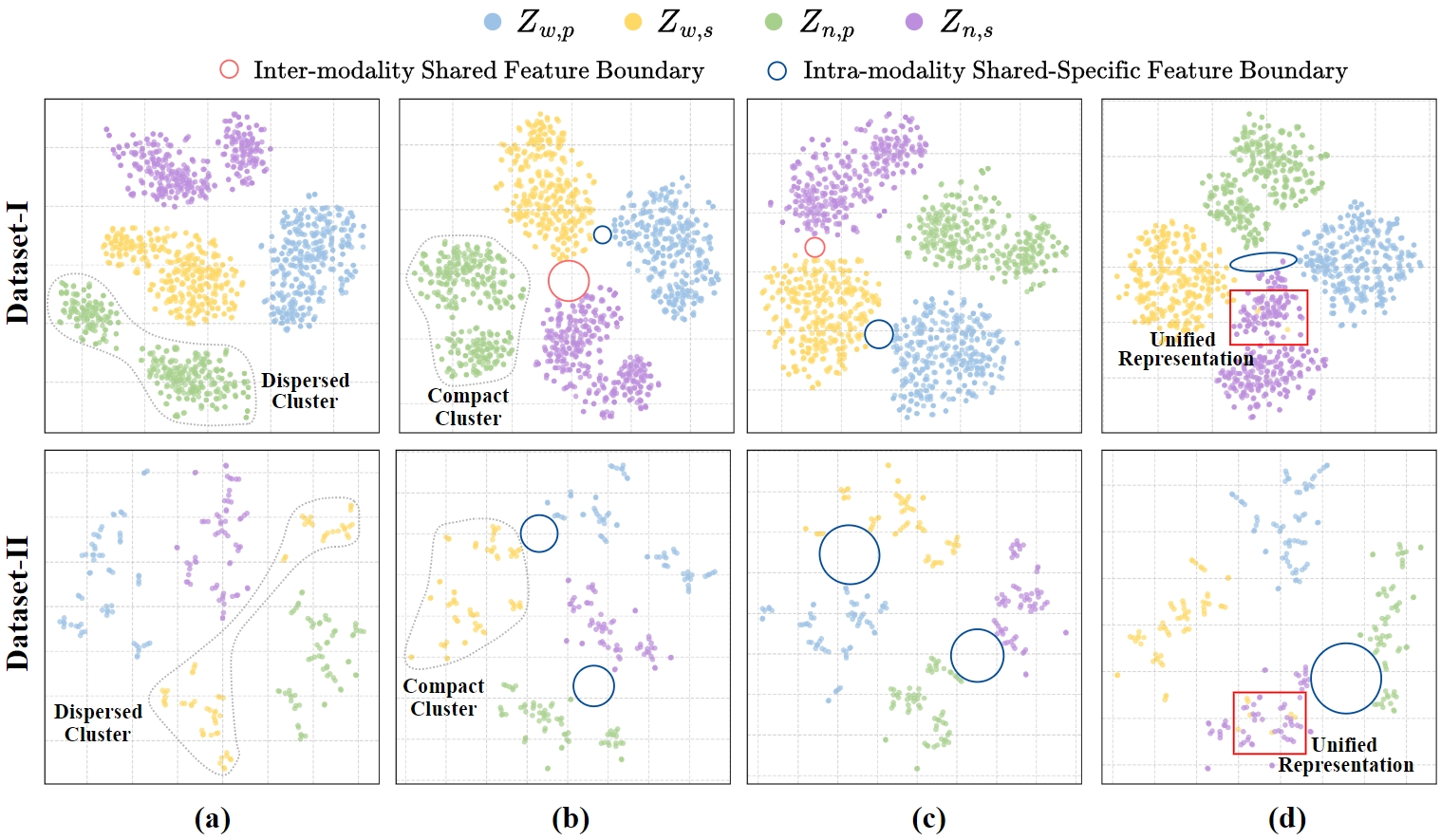}
    \caption{t-SNE visualization of ablation results: 
(a) Baseline, 
(b) with Distribution Alignment (DA),  
(c) with DA and Preliminary Disentanglement (PD), 
and (d) with DA, PD, and Disentangle-aware Contrastive Learning (DACL).}
    \label{fig:tsne}
\end{figure*}

\begin{table}[htbp]
\centering
\caption{Quantitative comparison with state-of-the-art methods on \textbf{Dataset-III}.}
\label{tab:dataset3_comparison}
\resizebox{0.5\textwidth}{!}{%
\begin{tabular}{c|cc|cccc} 
\toprule
\multicolumn{1}{c}{\multirow{2}[3]{*}{\textbf{Methods}}} & \multicolumn{2}{|c|}{\textbf{Modality}} & \multicolumn{4}{c}{\textbf{Metrics}} \\
\cmidrule{2-7}
& \textbf{WLI} & \textbf{NBI} & \textbf{IoU} & \textbf{Dice} & \textbf{SE} & \textbf{G-mean} \\ 
\midrule
TransUNet (2021)~\citep{chen2021transunet}         & \checkmark &            & 0.5322 & 0.6304 & 0.8532 & 0.9139 \\
TransUNet (2021)~\citep{chen2021transunet}        &            & \checkmark & 0.2313 & 0.3297 & 0.5042 & 0.6803 \\
\midrule
ShapeConv (2021)~\citep{ShapeConv}     & \checkmark & \checkmark & 0.5728 & 0.6791 & 0.7402 & 0.8514 \\
MTF (2022)~\citep{wang2022multimodaltokenfusionvision} & \checkmark & \checkmark & 0.5371 & 0.6479 & 0.7348 & 0.8467 \\
CMFNet (2022)~\citep{ma2022cmfnet}     & \checkmark & \checkmark & 0.5377 & 0.6263 & 0.6668 & 0.8114 \\
RS3Mamba (2024)~\citep{ma2024rs3mamba} & \checkmark & \checkmark & 0.5856 & 0.6734 & 0.7120 & 0.8386 \\
FTransUNet (2024)~\citep{ma2024fusion} & \checkmark & \checkmark & 0.5614 & 0.6636 & 0.7779 & 0.8698 \\
Dformer (2024)~\citep{yin2023dformer}  & \checkmark & \checkmark & 0.5978 & 0.7040 & 0.7829 & 0.8769 \\
\textbf{Ours}                         & \checkmark & \checkmark & \textbf{0.6362} & \textbf{0.7364} & \textbf{0.9174} & \textbf{0.9463} \\ 
\bottomrule
\end{tabular}
}
\end{table}

\begin{table*}[t]
\centering
\caption{Quantitative results of the ablation study.}
\label{tab:ablation}
\tiny
\resizebox{1\textwidth}{!}{ 
\begin{tabular}{cccc|cccc|cccc} 
\toprule
\multirow{2}[3]{*}{\textbf{DA}} & \multirow{2}[3]{*}{\textbf{PD}} & \multirow{2}[3]{*}{\textbf{DACL}} & \multirow{2}[3]{*}{\textbf{TS}} & \multicolumn{4}{c|}{\textbf{Dataset-I}} & \multicolumn{4}{c}{\textbf{Dataset-II}} \\ 
\cmidrule(lr){5-8} \cmidrule(lr){9-12}
                                        &                                                &                                                &                                                 & \textbf{IoU}        & \textbf{Dice}       & \textbf{SE}          & \textbf{G-mean}       & \textbf{IoU}        & \textbf{Dice}       & \textbf{SE}          & \textbf{G-mean}        \\ 
\midrule
\ding{55}                               & \ding{55}                                     & \ding{55}                                     & \ding{55}                                         & 0.6126              & 0.7158              & 0.8454              & 0.9013              & 0.7742              & 0.8530              & 0.9059              & 0.9498              \\
\ding{51}                               & \ding{55}                                     & \ding{55}                                     & \ding{55}                                         & 0.6260              & 0.7331              & 0.8510              & 0.9066              & 0.7979              & 0.8727              & 0.9053              & 0.9491              \\
\ding{51}                               & \ding{51}                                     & \ding{55}                                     & \ding{55}                                         & 0.6348              & 0.7406              & 0.8552              & 0.9075              & 0.8027              & 0.8750              & 0.9162              & 0.9554              \\
\ding{51}                               & \ding{51}                                     & \ding{51}                                     & \ding{55}                                         & 0.6402              & 0.7498              & 0.8539              & 0.9088              & 0.8129              & 0.8811              & 0.9165              & 0.9559              \\
\ding{51}                               & \ding{51}                                     & \ding{51}                                     & \ding{51}                                         & \textbf{0.6453}     & \textbf{0.7517}     & \textbf{0.8560}     & \textbf{0.9100}     & \textbf{0.8205}     & \textbf{0.8895}     & \textbf{0.9237}     & \textbf{0.9599}     \\ 
\bottomrule
\end{tabular}
}
\end{table*}

\subsection{Ablation Studies}

To assess the contributions of individual components within our proposed framework, we conducted extensive ablation studies. Each experiment selectively activates or deactivates specific modules—distribution alignment (DA), preliminary feature disentanglement (PD), and disentangle-aware contrastive learning (DACL)—to examine their respective impact on segmentation performance. The results, as summarized in Table~\ref{tab:ablation}, demonstrate the effectiveness of each component in enhancing segmentation accuracy across multiple datasets.

\textbf{Effectiveness of Distribution Alignment}  
We introduce distribution alignment (DA) to enhance global feature consistency across modalities. As shown in the experimental results, enabling DA alone significantly improves segmentation performance. For instance, on \textbf{Dataset-I}, the IoU increases from 0.6126 to 0.626 and the Dice score improves from 0.7158 to 0.7331. Similar improvements are observed on \textbf{Dataset-II}. These results underscore the importance of aligning multimodal feature distributions to fully exploit complementary information. Furthermore, as illustrated by the t-SNE visualizations in Figure~\ref{fig:tsne}, the inclusion of DA results in more compact and well-clustered feature distributions, in contrast to the scattered feature embeddings observed in its absence.

\textbf{Effectiveness of Preliminary Feature Disentanglement}  
Incorporating preliminary feature disentanglement (PD) further improves performance. When PD is applied together with DA, the IoU on Dataset-I increases from 0.626 to 0.641, and the Dice score rises from 0.733 to 0.748. Similar gains are observed on Dataset-II. PD facilitates the separation of shared and modality-specific features, improving the model's robustness in scenarios where distribution alignment alone may not suffice. As demonstrated by the t-SNE plots in Figure~\ref{fig:tsne}, the addition of PD results in semantically aligned shared features that are spatially adjacent, transitioning from previously uncorrelated distributions. Moreover, clear gaps emerge between shared and specific features within each modality, as well as between the specific features of different modalities, validating the effectiveness of disentanglement in learning structured representations.

\textbf{Effectiveness of Disentangle-aware Contrastive Learning}  
Finally, we evaluate the contribution of the disentangle-aware contrastive learning (DACL) component. When DACL is integrated with both DA and PD, the model achieves the best performance across all datasets. In addition, the t-SNE visualization in Figure~\ref{fig:tsne} reveals that DACL encourages deeper fusion of semantically similar shared features across modalities, resulting in more compact embeddings, while simultaneously enhancing the separation (\emph{i.e.}, margin) between dissimilar features. This indicates that our proposed progressive feature disentanglement learning strategy gradually refines and corrects cross-modal shared and private semantic features for segmentation tasks through the designed DA, PD, and DACL combination.

\textbf{Effectiveness of Progressive Training Strategy}  
We further validate the effectiveness of our progressive training strategy. By gradually increasing the weight of the feature disentanglement loss throughout training, the model is guided toward stable convergence and better representation learning. Experimental results on \textbf{Dataset-I} demonstrate that incorporating this progressive scheme leads to a performance improvement, with the IoU increasing from 0.6348 to 0.6453. This indicates that progressive weighting helps the model avoid premature overfitting to disentanglement constraints and improves overall segmentation accuracy.

In summary, each component in our framework—DA, PD, and DACL—plays a critical role in boosting segmentation performance. Their synergistic integration leads to substantial improvements across various benchmarks. Furthermore, our qualitative visualizations support these findings, illustrating that our full model reduces misidentification and produces more accurate and consistent predictions.

\subsection{Hyperparameter Analysis}
To ensure effective feature disentanglement, we conducted controlled experiments to investigate the impact of different weighting schemes for the subcomponents within the disentanglement loss. Specifically, to maintain numerical stability and avoid disproportionate loss scaling, we define the weighting parameters as follows.
 
Given that the auxiliary losses—\(\mathcal{L}_{\text{Align}}, \mathcal{L}_{\text{Diff}}, \mathcal{L}_{\text{Orth}}\)—are all normalized within the range \([0, 1]\), directly adding them without adjustment could lead to overly dominant gradients. To avoid this, we assign them equal contributions by setting the weights \(\alpha = \beta = \gamma = \frac{1}{3}\), thereby ensuring a balanced influence in the total disentanglement objective.

In contrast, the disentangle-aware contrastive loss \(\mathcal{L}_{\text{DACL}}\) typically produces larger positive values and poses greater challenges in convergence. If left unregulated, it may dominate the optimization process and hinder the learning of other objectives. To mitigate this issue, we introduced a separate scaling factor \(\delta\) to control its relative contribution. Through empirical testing with \(\delta\) ranging from $1$ to $0.001$, we found that setting \(\delta = 0.01\) achieves the best trade-off. This value ensures effective disentanglement without overwhelming the segmentation losses, ultimately leading to stable training and improved performance.

\begin{table*}[t]
\centering
\caption{Ablation study results with different parameter settings.}
\label{tab:param}
\resizebox{1\textwidth}{!}{ 
\begin{tabular}{cccc|cccc|cccc} 
\toprule
\multirow{2}{*}{\boldmath$\alpha$} & \multirow{2}{*}{\boldmath$\beta$} & \multirow{2}{*}{\boldmath$\gamma$} & \multirow{2}{*}{\boldmath$\delta$} & \multicolumn{4}{c|}{\textbf{Dataset-I}} & \multicolumn{4}{c}{\textbf{Dataset-II}} \\ 
\cmidrule(lr){5-8} \cmidrule(lr){9-12}
                                        &                                                &                                                &                                                 & \textbf{IoU}        & \textbf{Dice}       & \textbf{SE}          & \textbf{G-mean}       & \textbf{IoU}        & \textbf{Dice}       & \textbf{SE}          & \textbf{G-mean}        \\ 
\midrule
1.000 & 1.000 & 1.000 & 1.000 & 0.6123 & 0.7151 & 0.8578 & 0.9085 & 0.7552 & 0.8345 & 0.9147 & 0.9546 \\
0.333 & 0.333 & 0.333 & 0.100 & 0.6245 & 0.7290 & 0.8585 & 0.9084 & 0.8097 & 0.8805 & 0.9253 & 0.9605 \\
0.333 & 0.333 & 0.333 & 0.010 & \textbf{0.6453} & \textbf{0.7517} & \textbf{0.8560} & \textbf{0.9100} & \textbf{0.8205} & \textbf{0.8895} & \textbf{0.9237} & \textbf{0.9599} \\
0.333 & 0.333 & 0.333 & 0.001 & 0.6243 & 0.7310 & 0.8486 & 0.9054 & 0.8097 & 0.8805 & 0.9068 & 0.9508 \\

\bottomrule
\end{tabular}
}

\end{table*}

\section{Conclusion}
In this work, we proposed a multimodal learning framework for 2D medical endoscopic image analysis, integrating White Light Imaging and Narrow Band Imaging. By employing multi-scale distribution alignment and progressive feature disentanglement techniques, we effectively addressed modality discrepancies and facilitated semantic fusion. Extensive experiments across multiple datasets validated the superior accuracy and generalization of our method, showcasing its potential for advancing automated medical image segmentation.

\section*{Acknowledgments}
This paper was supported by the National Key R\&D Program of China under Grant 2022YFA1008300, National Natural Science Foundation of China under Grants 12471308, Guangdong Provincial Science and Technology Plan under Grant 2022B1515130009.

{\bibliographystyle{cas-model2-names}
\bibliography{egbib}
}

\end{document}